\documentclass[twocolumn,showpacs,preprintnumbers,amsmath,amssymb]{revtex4}
\usepackage{graphicx}  
\usepackage{dcolumn}   
\usepackage{bm}        
\usepackage{amssymb}   
\usepackage[colorlinks,citecolor=black,linkcolor=black,hyperindex,pdfborder={0 0 1},dvipdfm]{hyperref}%

\begin{document}
\title{Non-Markovian dynamics without using quantum trajectory}

\author{Chengjun Wu}
\author{Yang Li}
\author{Mingyi Zhu}
\author{Hong Guo}
\email[Correspondence author:\ ]{hongguo@pku.edu.cn}

\affiliation{CREAM Group, State Key Laboratory of Advanced Optical
Communication Systems and Networks (Peking University)
\\School of Electronics Engineering
and Computer Science, Peking University, Beijing 100871, China}


\date{\today}

\begin{abstract}Open quantum system interacting with structured environment is important and manifests non-Markovian behavior,
which was conventionally studied using quantum trajectory stochastic
method. In this paper, by dividing the effects of the environment
into two parts, we propose a deterministic method without using
quantum trajectory. This method is more efficient and accurate than
stochastic method in most Markovian and non-Markovian cases. We also
extend this method to the generalized Lindblad master equation.
\end{abstract}


\pacs{03.65.Yz, 42.50.Lc}


\maketitle


When an open quantum system interacts with environment, it
experiences decoherence and dissipation which lead to loss of
information. Such open quantum system is depicted by a reduced
density matrix which shows non-unitary evolution. On the other hand,
the environment is classified as Markovian with no memory effect,
and non-Markovian with memory effect. In Markovian case, since there
is no memory effect, the quantum trajectory based Monte Carlo wave
function (MCWF) method \cite{QJ1, QJ3, MCM} and quantum state
diffusion (QSD) method \cite{QSDM1, QSDM2} are applied. However, in
non-Markovian case, due to memory effect, the information lost by
the system during the interaction with the environment will come
back to the system in a later time and so shows much more
complicated behaviors than Markovian case.

Non-Markovian systems are important for their applications to many
fields of physics, such as quantum information processing
\cite{quantum information, LiZhou}, quantum optics \cite{OQS}, solid
state physics \cite{solid state physics}, and chemical physics
\cite{chemical physics}. Recently, non-Markovian behaviors have also
been studied in biomolecules where the molecules are embedded in a
solvent and/or in a protein environment \cite{biomolecules}. Since
there is no true pure state quantum trajectory due to the memory
effect \cite{no single trajectory}, the quantum trajectory based
Markovian methods do not work. Thus, doubled Hilbert space (DHS)
method \cite{DHS}, triple Hilbert space (THS) method \cite{THS},
non-Markovian QSD method \cite{QSD1, QSD2}, and non-Markovian
quantum jump (NMQJ) method \cite{NMQJ1, NMQJ2} are proposed to solve
the non-Markovian dynamics of the system where the memory effect is
taken into account. However, in order to obtain high accuracy, all
these methods, which are based on stochastic simulations, need to
fulfill a large number of realizations and is very time-consuming.
So, new methods which are more efficient and accurate are highly
desired.

In this paper, a deterministic method without using quantum
trajectory is proposed to solve the non-Markovian dynamics. The
influence of the environment on the system is divided into two
parts, i.e., the non-unitary evolution of the states and the
probability flow between these states. Moreover, we also extend this
approach to the generalized Lindblad master equation which can deal
with some strong coupling cases \cite{GLE}. The algorithm and
numerical efficiency are given, which show that our method is more
efficient and accurate than those based on stochastic simulation in
most Markovian and non-Markovian cases.

The dynamics of the non-Markovian system is governed by the
following master equation \cite{OQS}
\begin{equation}\label{eqn1}
\begin{split}
 \dot \rho (t) = &\displaystyle\frac{1}{{i\hbar }}[H_s ,\rho (t)] + \sum\limits_j {\gamma _j (t)C_j (t)\rho (t)C_j^\dag  (t)}  \\
 &- \displaystyle\frac{1}{2}\sum\limits_j {\gamma _j (t)\{ \rho (t),C_j^\dag  (t)C_j (t)\} } , \\
\end{split}
\end{equation}
where $H_s$ is the system Hamiltonian including the Lamb shift,
$C_j(t)$ are the jump operators which induce changes [e.g., jump
from state $\psi_{\alpha} (t)$ to $\psi_{\alpha'} (t)$ i.e.,
$|\psi_{\alpha'}(t)\rangle = C_j(t)
|\psi_{\alpha}(t)\rangle/\left|\left| C_j(t)
|\psi_{\alpha}(t)\rangle\right|\right|$ ] in the system, and
$\gamma_j(t)$ are the decay rates which may take negative values for
some time intervals. The reduced density matrix can be written as
\cite{NMQJ1}
\begin{equation}\label{eqn2}
\rho(t) = \sum_{\alpha=1}^{N_{eff}} p_{\alpha} (t)
|\psi_{\alpha}(t)\rangle\langle \psi_{\alpha}(t)|,
\end{equation}
where $p_{\alpha} (t)$ is the probability of the system being in the
state $|\psi_{\alpha}(t)\rangle$ at time $t$. Further, it should be
pointed out that the effective number of the states $N_{eff}$ is
determined by $C_j (t)$'s \cite{NMQJ2}, $\sum_{\alpha = 1}^{N_{eff}}
p_{\alpha} (t) = 1$ and that the state $| \psi_{\alpha} (t)\rangle$
is normalized.

To solve the dynamics of the system, one should know the time
evolution of $|\psi_{\alpha}(t)\rangle$ and its probability
$p_{\alpha} (t)$. In our method, the time evolution of the state
$|\psi_{\alpha}(t)\rangle$ is the same as that in NMQJ \cite{NMQJ1}.
In NMQJ, the probability $p_{\alpha} (t)$ is calculated in a
stochastic way by using quantum trajectory to $N$ ensemble members.
In our method, however, the evolution of probability $p_{\alpha}
(t)$ is given in a deterministic way:
\begin{equation}\label{eqn3}
\dot p_{\alpha} (t) =  - \sum_j \Gamma_{\alpha}^j (t) p_{\alpha} (t)
+ {\sum_{(\alpha', j)}}'\Gamma_{\alpha'}^j (t) p_{\alpha'} (t),
\end{equation}
where $\Gamma_{\alpha} ^j (t) = {\gamma _j (t)}{\left\| {C_j (t)\left| {\psi _{\alpha } (t)} \right\rangle } \right\|}^2$ and ${\sum\limits_{(\alpha', j)}}'$
represents the summation over all the pairs $(\alpha', j)$
satisfying $\left| {\psi _\alpha  (t)} \right\rangle  = {{C_j
(t)\left| {\psi _{\alpha '} (t)} \right\rangle } \mathord{\left/
 {\vphantom {{C_j (t)\left| {\psi _{\alpha '} (t)} \right\rangle } {\left\| {C_j (t)\left| {\psi _{\alpha '} (t)} \right\rangle } \right\|}}} \right.
 \kern-\nulldelimiterspace} {\left\| {C_j (t)\left| {\psi _{\alpha '} (t)} \right\rangle } \right\|}}
$. One finds that the probability of the state, $p_{\alpha} (t)$,
changes via the mechanism of jumps for ``out" ($\alpha \to \alpha'$)
and ``in" ($\alpha' \to \alpha$), respectively.

The numerical simulation corresponding to Eq. (\ref{eqn3}) is
straightforward:
\begin{equation}\label{eqn4}
p_{\alpha} (t + \delta t) = p_{\alpha} (t) - \delta t  \sum_j
\Gamma_{\alpha}^j (t) p_{\alpha} (t) + \delta t  {\sum_{(\alpha',
j)}}'\Gamma_{\alpha'}^j (t) p_{\alpha'} (t).
\end{equation}
Note that there is no stochastic noise and no need to consider the
sign of the decay rate during the simulation. Additionally, the
$p_{\alpha}(t)$'s in our method do represent the probability of the
system actually being in the corresponding pure state ensemble.

Consider a particular transition: $|\psi_{\alpha'}(t)\rangle =
C_j(t) |\psi_{\alpha}(t)\rangle/\left|\left| C_j(t)
|\psi_{\alpha}(t)\rangle\right|\right|$, then the corresponding
probability change takes the form:
\begin{equation}\label{eqn5}
\begin{split}
p_\alpha  (t+ \delta t) &=p_\alpha  (t) -\delta t p_\alpha(t)\Gamma_\alpha ^j (t),\\
p_{\alpha '}   (t+ \delta t)& =p_{\alpha '} (t) +\delta t p_\alpha
(t)\Gamma_\alpha ^j (t).\\
\end{split}
\end{equation}
When the decay rate $ \gamma_j (t) $ is positive or negative, the
probability flow is from $ \left| {\psi _\alpha (t)} \right\rangle $
to $|\psi_{\alpha'}(t)\rangle $ or reversed. This has been mentioned
in Ref.\cite{NMQJ1}. However, it is more explicit in our method.
From Eq. (\ref{eqn5}), it is clear that, in the negative decay
region, the amount of probability flow only depends on the target
state and the probability of the system being in the target state.
This is similar to the situation in NMQJ \cite{NMQJ1}, where the
jump probability in the negative decay region is proportional to the
number of particles in the target state. These indicate that the
trajectory of a particle in NMQJ can not be interpreted as true
trajectory since the jump process depends on the status of other
particles in the system. Because true pure state quantum
trajectories do not exist in the non-Markovian dynamics \cite{no
single trajectory}, it is not necessary to calculate $p_{\alpha}(t)$
in a stochastic way.


Next, we extend our method to the recently proposed generalized
Lindblad master equation which can solve the dynamics of some highly
non-Markovian systems\cite{GLE},
\begin{equation}\label{eqn6}
\frac{d}{{dt}}{\rho _i} =  - i[{H_i},{\rho _i}] +
\sum\limits_{j\lambda } {\left( {R_\lambda ^{ij}{\rho _j}R{{_\lambda
^{ij}}^\dag } - \frac{1}{2}\{ R{{_\lambda ^{ji}}^\dag }R_\lambda
^{ji},{\rho _i}\} } \right)} ,
\end{equation}
where $i , j=1,2,\cdots,n$, $H_i$ are any Hermitian
operators, and $R_{\lambda}^{ij}$ are any system operators. 
It should be indicated that $\rho (t) = \sum\limits_{i = 1}^n {{\rho
_i}(t)}$.

The $i$th density matrix is decomposed as:
\begin{equation}\label{eqn7}
{ \rho_i}(t) = \sum\limits_{\alpha  = 1}^{N_{eff}^i} {p_i^\alpha
(t)\left| {\psi _i^\alpha (t)} \right\rangle \left\langle {\psi
_i^\alpha (t)} \right|},
\end{equation}
where $N_{eff}^i$ is determined in the same way in Eq. (\ref{eqn2})
by taking all the jump operators ${R_\nu ^{ij}}^,$s and all the
states $\psi _j^\alpha (t)^,$s in each ${ \rho_j}(t)$ into
consideration.

The evolution of state $\left| {\psi _i^\alpha (t)} \right\rangle$
is governed by the nonlinear differential equation\cite{example2}
\begin{equation}\label{eqn8}
i\frac{d}{{dt}}\left| {\psi _i^\alpha (t)} \right\rangle  = \hat G
(\psi _i^\alpha )(t)\left| {\psi _i^\alpha (t)} \right\rangle ,
\end{equation}
where $\hat G(\psi _i^\alpha )(t) = {H^i} -
\frac{i}{2}\sum\limits_{j\nu } {R{{_\nu ^{ji}}^\dag }R_\nu ^{ji}}  +
\frac{i}{2}\sum\limits_{j\nu } {{{\left\| {R_\nu ^{ji}\left| {\psi
_i^\alpha (t)} \right\rangle } \right\|}^2}} $. By combining Eqs.
(\ref{eqn6}), (\ref{eqn7}), (\ref{eqn8}) and noting that $ {\left|
{\psi _i^1 (t)} \right\rangle \left\langle {\psi _i^1 (t)} \right|}
$, $ {\left| {\psi _i^2 (t)} \right\rangle \left\langle {\psi _i^2
(t)} \right|} $, $ \cdots $,  $ {\left| {\psi _i^{N_{eff}^i} (t)}
\right\rangle \left\langle {\psi _i^{N_{eff}^i} (t)} \right|} $ are
linearly independent, the evolution of $p_i^{\alpha} (t)$ is given
by
\begin{equation}\label{eqn9}
\dot p_i^\alpha (t) =  - \sum\limits_{j\nu } {\Gamma _{\nu \alpha
}^{ji}} p_i^\alpha (t) + {\sum\limits_{(j,\nu ,\alpha ')}}' {\Gamma
_{\nu \alpha '}^{ij}p_j^{\alpha '}(t)} ,
\end{equation}
where $\Gamma _{\nu \alpha }^{ij} = {\left\| {R_\nu ^{ij}\left|
{\psi _j^\alpha (t)} \right\rangle } \right\|^2}$ and
${\sum\limits_{(j,\nu ,\alpha ')}}'$ represents the summation over
all the pairs $(j,\nu ,\alpha ')$ satisfying $\left| {\psi _i^\alpha
(t)} \right\rangle  = {{R_\nu ^{ij}\left| {\psi _j^{\alpha '}(t)}
\right\rangle } \mathord{\left/
 {\vphantom {{R_\nu ^{ij}\left| {\psi _j^{\alpha '}(t)} \right\rangle } {\left\| {R_\nu ^{ij}\left| {\psi _j^{\alpha '}(t)} \right\rangle } \right\|}}} \right.
 \kern-\nulldelimiterspace} {\left\| {R_\nu ^{ij}\left| {\psi _j^{\alpha '}(t)} \right\rangle } \right\|}}$.
It can be easily seen that by setting $n=1$ and taking the decay
rates $\gamma (t)$ into the equation, Eq. (\ref{eqn9}) degenerates
to Eq. (\ref{eqn3}).

\emph{Example $1$: Detuned Jaynes-Cummings model.}--Consider a
system with a two-level atom in a detuned damped cavity, which is
governed by the time convolutionless master equation \cite{OQS}
\begin{equation}\label{eqn10}
\begin{split}
 \dot\rho (t)=  &- \displaystyle\frac{i}{2}S(t)\{ \sigma _ +  \sigma _ -  ,\rho  (t)\}  \\
 &  + \gamma (t)\{ \sigma _ -  \rho (t)\sigma _ +   - \displaystyle\frac{1}{2}\sigma _ +  \sigma _ -  \rho  (t) - \displaystyle\frac{1}{2}\rho  (t)\sigma _ +  \sigma _ -  \} . \\
 \end{split}
\end{equation}
The spectral density of the cavity is supposed to be of Lorentzian
profile, i.e., $J(\omega ) =\frac{\gamma _0 \lambda ^2
}{2\pi[(\omega _0 - \Delta - \omega) ^2  + \lambda ^2 ]},$ where $
\Delta = \omega_0 - \omega_c$ is the detuning between the cavity
mode and the atom. To second order approximation, the Lamb shift and
the decay rate take the form \cite{OQS} $S(t) = \frac{{\gamma _0
\lambda \Delta }}{{\lambda ^2 + \Delta ^2 }}\{1 - {\mathop{\rm
e}\nolimits} ^{ - \lambda t} [\cos (\Delta t) + \frac{\lambda
}{\Delta }\sin (\Delta t)]\},$ $\gamma (t) = \frac{{\gamma _0
\lambda ^2 }}{{\lambda ^2 + \Delta ^2 }}\{1 - {\mathop{\rm
e}\nolimits} ^{ - \lambda t} [\cos (\Delta t) - \frac{\Delta
}{\lambda }\sin (\Delta t)]\}.$ In this model, there is only one
jump operator $ C = \sigma _ - = \left| g \right\rangle \left\langle
e \right| $, which is a lowering operator. We assume that $\rho (0)
= \left| {\psi _1(0) } \right\rangle \left\langle {\psi _1(0) }
\right| $ and choose $ \left| {\psi _1 (0)} \right\rangle  =
(4\left| e \right\rangle  + 3\left| g \right\rangle )/5 $. Acting
the jump operator on the state $ \left| {\psi _1(0) } \right\rangle
$, we get $ \left| {\psi _2 (0)} \right\rangle  = \left| g
\right\rangle $. According to Eq. (\ref{eqn4}), at time $ t + \delta
t $, the probabilities become
\begin{equation}\label{eqn11}
\begin{array}{l}
 p_1 (t + \delta t) = p_1 (t) - \delta tp_1 (t)\Gamma_1^1 (t), \\
 p_2 (t + \delta t) = p_2 (t) + \delta tp_1 (t)\Gamma_1^1 (t), \\
 \end{array}
\end{equation}
where $\Gamma_1^1 (t) = \gamma (t)\left| {\left\langle {e}
 \mathrel{\left | {\vphantom {e {\psi _1 (t)}}}
 \right. \kern-\nulldelimiterspace}
 {{\psi _1 (t)}} \right\rangle } \right|^2.$

In this example, $\rho_{ee}(t)$ is proportional to the energy of the
system and $p_2 (t)$ represents the probability for one photon being
in the environment. Although $p_1(t)$ and $p_2(t)$ can be solved
analytically, in order to illustrate our method, we use Eq.
(\ref{eqn11}) to do the simulation. The parameters are chosen as
$\Delta = 12\lambda,
 \gamma_0\lambda  = 4, \lambda \delta t=0.005$.
\begin{figure}[htbp]
\centering
\includegraphics[width=3.3in]{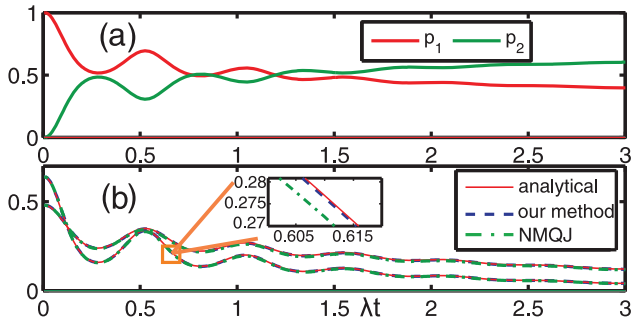}
\caption{\small (color online) Dynamics of detuned Jaynes-Cummings
model. The initial state is $ \left| {\psi _1 (0)} \right\rangle =
(4\left| e \right\rangle  + 3\left| g \right\rangle )/5 $ and the
parameters are $\Delta = 12\lambda,
 \gamma \lambda = 4, \lambda \delta t=0.005$. (a) The probabilities for the system in states $ \left| {\psi _1 (t)}
\right\rangle $ and $ \left| {\psi _2 (t)} \right\rangle $. (b) The
population of the excited state $\rho_{ee}$ (initially higher line)
and the absolute value of the coherence $\rho_{eg}$ (initially lower
line) with three methods: analytic (red solid curve), our method
(blue long-dashed curve) and NMQJ (with $N=10^4$
particles in the system, green dash-dot curve). 
}
\label{pic1}
\end{figure}
Figure \ref{pic1} (a) shows explicitly the reversal of the
probability flow. We can see from Fig. \ref{pic1} (a) and (b) that
when the probability flow gets reversed, the energy and coherence of
the atom increase. These show
explicitly the memory effect that the reduced system restores the information lost earlier. 
In Fig. \ref{pic1} (b), the result of NMQJ (with $N=10^4$ particles
in the system) is also given, which shows that our method is more
accurate.

\emph{Example 2: Application to generalized Lindblad master
equation.}-- To illustrate our method for this kind of equation, we
consider a two-state system coupled to an environment consisting of
two energy bands, each with a finite number of evenly spaced levels.
This may be viewed as a spin coupled to a single molecule or a
single particle quantum dot \cite{example21}. By using
time-convolutionless projection operator technique, to the second
order, the generalized Lindblad master equation takes the form
\cite{example2}
\begin{equation}\label{eqn12}
\begin{array}{l}
 \displaystyle\frac{d}{{dt}}\rho _1  = \int_0^t {dt_1 h(t - t_1 )} [2\gamma _1 \sigma ^ +  \rho _2 \sigma ^ -   - \gamma _2 \{ \sigma ^ +  \sigma ^ -  ,\rho _1 \} ], \\
 \displaystyle \frac{d}{{dt}}\rho _2  = \int_0^t {dt_1 h(t - t_1 )} [2\gamma _2 \sigma ^ -  \rho _1 \sigma ^ +   - \gamma _1 \{ \sigma ^ -  \sigma ^ +  ,\rho _2 \} ], \\
 \end{array}
\end{equation}
where $\gamma_i h(t-t_1), (i=1,2),$ is the environment correlation
function with $h(t)=\frac{{\delta \varepsilon \sin ^2 (\delta
\varepsilon t/2)}}{{2\pi (\delta \varepsilon t/2)^2 }}$ where
$\delta \varepsilon$ is the width of the upper and lower energy
bands. The reduced density matrix for the system is given by $ \rho
= \rho _1  + \rho _2.$
\begin{figure}[h]
\centering
\includegraphics[width=3.3in]{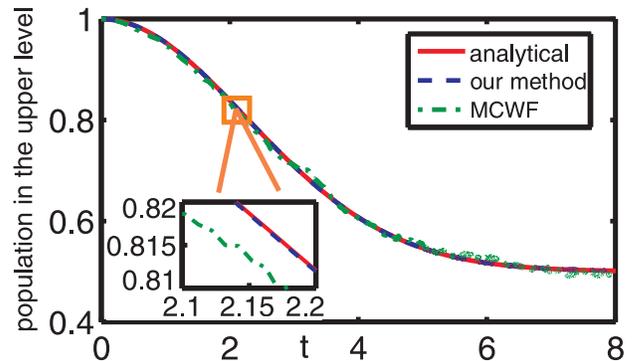}
\caption{\small (color online) A two-state system coupled to an
environment consisting of two energy bands. Comparison of our method
(blue long-dashed curve) and Monte Carlo simulation (with $N=10^4$
trajectories, green dash-dot curve) to analytical result (red solid
curve). The parameters are $\delta \epsilon =0.31,$
$\gamma_1=\gamma_2=1$ and time step $\delta t=0.01$. } \label{pic3}
\end{figure}
We assume that $ \rho _1 (0) = \left| e \right\rangle \left\langle e
\right| $ and $\rho_2(0)=0$. The parameters are chosen as $\delta
\epsilon =0.31$ and $\gamma_1=\gamma_2=1$. In Fig. \ref{pic3} we
compare the results of our method, analytical solution and Monte
Carlo simulation which is based on the unraveling of the master
equation (with $N=10^4$ trajectories) \cite{example2}. Apparently,
our method is more accurate than Monte Carlo simulation method.

According to Eqs. (\ref{eqn4}), we only need to calculate $N_{eff}$
states and change the probabilities deterministically. The time cost
is almost determined by the calculation of $N_{eff}$ states.
However, the evolution of $ N_{eff}$ states is independent with each
other, so we can calculate them parallelly. In addition, if the jump
operators can be represented by sparse matrixes, we only need to
calculate the evolution of the states appearing in the decomposition
of $\rho (0)$ and use the jump operators to obtain other states.
Moreover, since the sign of the decay rate makes no difference
during the simulation, in non-Markovian case, our method is as
efficient as it behaves in Markovian case.

Similar to  our method, the NMQJ method \cite{NMQJ1, NMQJ2} needs to
calculate $ N_{eff}$ states. However, in addition to that, NMQJ has
to consider the sign of the decay rates and generate $N$ random
numbers ($N \gg N_{eff}$) to decide the jump process at each time
step $\delta t$. Apparently, our method is more efficient than NMQJ
in any case.

In Markovian case, the MCWF \cite{QJ1} and QSD \cite{QSDM1}  method
need to realize a large number of trajectories for every state
appearing in the decomposition of $\rho(0)$. When the number of
these trajectories is larger than $ N_{eff}$, which is always the
case, our method is more efficient than them. In non-Markovian case,
the DHS method \cite{DHS}, THS method \cite{THS} and non-Markovian
QSD method \cite{QSD1, QSD2} all introduce additional cost for
computational efficiency compared to MCWF or QSD. However, in
non-Markovian case, our method is as efficient as it behaves in
Markovian case. Thus, when the number of these trajectories is
larger than $ N_{eff}$, our method is obviously more efficient than
them, too.

As for the accuracy, since there is no statistical noise in our
method and the error caused by finite time step $\delta t$ is the
same, compared with all the methods based on stochastic simulation,
our method is more accurate. Actually, our method is the limit case
when the number of realizations in the stochastic based methods
tends to infinite.

In conclusion, by dividing the influence of the environment on the
system into two parts, i.e., the non-unitary evolution of these
states and the probability flow between them, we propose a
deterministic method to solve the non-Makovian dynamics. Compared
with the method based on stochastic simulation, our method has
advantages in efficiency and accuracy. Additionally, we extended
this approach to the generalized Lindblad master equation , which is
useful to solve the dynamics of some highly non-Markovian systems.

This work is supported by the Key Project of the National Natural
Science Foundation of China (Grant No. 60837004).

\end{document}